\documentclass[conference]{IEEEtran}

\usepackage[pdftex]{graphicx}

\usepackage{scalerel}
\usepackage{tikz}
\usetikzlibrary{svg.path}
\definecolor{orcidlogocol}{HTML}{A6CE39}
\tikzset{
  orcidlogo/.pic={
    \fill[orcidlogocol] svg{M256,128c0,70.7-57.3,128-128,128C57.3,256,0,198.7,0,128C0,57.3,57.3,0,128,0C198.7,0,256,57.3,256,128z};
    \fill[white] svg{M86.3,186.2H70.9V79.1h15.4v48.4V186.2z}
                 svg{M108.9,79.1h41.6c39.6,0,57,28.3,57,53.6c0,27.5-21.5,53.6-56.8,53.6h-41.8V79.1z M124.3,172.4h24.5c34.9,0,42.9-26.5,42.9-39.7c0-21.5-13.7-39.7-43.7-39.7h-23.7V172.4z}
                 svg{M88.7,56.8c0,5.5-4.5,10.1-10.1,10.1c-5.6,0-10.1-4.6-10.1-10.1c0-5.6,4.5-10.1,10.1-10.1C84.2,46.7,88.7,51.3,88.7,56.8z};
  }
}
\newcommand\orcidicon[1]{\href{https://orcid.org/#1}{\mbox{\scalerel*{
\begin{tikzpicture}[yscale=-1,transform shape]
\pic{orcidlogo};
\end{tikzpicture}
}{|}}}}

\usepackage{url}
\usepackage[utf8]{inputenc}
\usepackage{textcomp}
\usepackage{gensymb}
\usepackage[T1]{fontenc}
\usepackage{hyperref}
\hypersetup{
	colorlinks,
	citecolor=black,
	filecolor=black,
	linkcolor=black,
	urlcolor=black
}

\definecolor{darkblue}{rgb}{0, 0, 0.5}

\usepackage{todonotes}

\makeatletter
\def\ps@IEEEtitlepagestyle{%
  \def\@oddfoot{\mycopyrightnotice}%
  \def\@evenfoot{}%
}
\def\mycopyrightnotice{%
  \begin{minipage}{\textwidth}
  \centering \scriptsize
  Copyright~\copyright~2024 IEEE.  Personal use of this material is permitted.  Permission from IEEE must be obtained for all other uses, in any current or future media, including\\reprinting/republishing this material for advertising or promotional purposes, creating new collective works, for resale or redistribution to servers or lists, or reuse of any copyrighted component of this work in other works.
  \end{minipage}
}
\makeatother

\begin{document}

% \title{Evolution of a Bundle Protocol Stack to a Software-Defined DTN Implementation}
% \title{µD3TN: Architectural Evolution Toward a Software-Defined DTN Implementation}
\title{The Architectural Refinement of µD3TN:\\Toward a Software-Defined DTN Protocol Stack}

% @all: Please adjust as necessary. -FW
\author{
\IEEEauthorblockN{Felix Walter\IEEEauthorrefmark{1}\orcidicon{0000-0002-4724-8092}, Marius Feldmann\IEEEauthorrefmark{1}, Juan A. Fraire\IEEEauthorrefmark{2}\orcidicon{0000-0001-9816-6989}, Scott Burleigh\IEEEauthorrefmark{2}}
\IEEEauthorblockA{\IEEEauthorrefmark{1}D3TN GmbH, Dresden, Germany}
\IEEEauthorblockA{\IEEEauthorrefmark{2}D3TN U.S. Corp., Miami, FL, USA}
\IEEEauthorblockA{<firstname>.<lastname>@d3tn.com}
    % @Juan: feel free to add your other affiliations you want to appear in the paper
}

\maketitle

%%%%%%%%%%%

\begin{abstract}
This paper provides a comprehensive overview of the µD3TN project's development, detailing its transformation into a flexible and modular software implementation of the Delay-/Disruption-Tolerant Networking (DTN) Bundle Protocol. 
Originating from µPCN, designed for microcontrollers, µD3TN has undergone significant architectural refinement to increase flexibility, compatibility, and performance across various DTN applications. 
Key developments include achieving platform independence, supporting multiple Bundle Protocol versions concurrently, introducing an abstract Convergence-Layer Adapter (CLA) interface, and developing the so called Application Agent Protocol (AAP) for interaction with the application layer. 
Additional enhancements, informed by field tests, include Bundle-in-Bundle Encapsulation and exploring a port to the Rust programming language, indicating the project's ongoing adaptation to practical needs. 
The paper also introduces the Generic Bundle Forwarding Interface and AAPv2, showcasing the latest innovations in the project. 
Moreover, it provides a comparison of µD3TN's architecture with the Interplanetary Overlay Network (ION) protocol stack, highlighting some general architectural principles at the foundation of DTN protocol implementations.
% maybe rather mention: highlighting the different target use cases of the implementations and giving a recommendation which one to select depending on the scenario

\end{abstract}

\IEEEpeerreviewmaketitle

%%%%%%%%%%%

\section{Introduction}

\subsection{Background}

% Intro to DTN and BP and CLA
Delay-Tolerant Networking (DTN) is increasingly pivotal in shaping the future of space networks.
% Motivation
As outlined in the Interagency Operations Advisory Group's (IOAG) report on Mars Communications Architecture~\cite{ioagreport}, the significance of DTN in space communications is increasingly recognized. 
Projects like NASA's Lunar Communications Relay and Navigation Systems (LCNRS)~\cite{TEMPOLCRNS}, the European Space Agency's (ESA) Moonlight initiative~\cite{giordano2021moonlight, ESAMoonlight}, and the collaborative effort for the LunaNet Interoperability Specification~\cite{israel2020lunanet, giordano2023lunanet, NASALunanet, NASA2023LunanetSpec} underscore this trend. 
This advancement mandates that nodes, whether located on planetary surfaces, in orbit or in space, implement DTN capabilities. 

% BP and CLA definitions
At the heart of DTN is the Bundle Protocol (BP)~\cite{rfc5050, rfc9171, CCSDS-Bundle}, designed to manage the transmission of Protocol Data Units, termed \emph{bundles}. 
These bundles are uniquely characterized by their ability to endure extended periods either stored in the memory of an intermediate node or in-transit across interplanetary distances, contrasting them with traditional Internet packets. 
DTN further incorporates convergence-layer adapters (CLAs). 
These integral components enable the integration of DTN into various underlying network architectures, ensuring seamless communication over diverse network types, including the Internet or deep-space links. 
%This sets the stage for a detailed exploration of various DTN protocol stack implementations, each offering distinct features and capabilities to address the complex demands of space communication.

% Existing DTN stacks:
\subsection{DTN/BP Stacks Overview}

Various actively-maintained protocol stacks enrich the landscape of DTN, each contributing unique features and capabilities to the field, addressing a range of requirements from space communications to more terrestrial applications.
\begin{itemize}
    % ION
    \item \textbf{ION}~\cite{ION-DTN}, developed by NASA's Jet Propulsion Laboratory (JPL), is a prominent implementation, offering extensive support for both BPv6 and BPv7, including security extensions (BPSec) and CLAs like TCPCLv3, UDPCL, and LTPv1. Its notable features include support for the Bundle Streaming Service Protocol (BSSP) and the Asynchronous Message Service (AMS).
    % IONE
    \item \textbf{IONe}~\cite{ione_sourceforge} is a fork of ION for exploring experimental features that are not yet flight-ready.
    %shares many similarities with ION but includes novel experimental features.
    % HDTN
    \item \textbf{HDTN}~\cite{HDTN}, developed by NASA's Glenn Research Center, is another significant stack. It distinguishes itself with Real-time Transport Protocol (RTP) support in BPv7 and a focus on high-performance networking. It also provides BPSec and BPv6 support.
    % DTNME
    \item \textbf{DTNME}~\cite{DTNME}, developed by NASA Marshall Space Flight Center, is a flexible DTN stack used for payload operations in the International Space Station (ISS).
    % CFS
    \item \textbf{CFS}~\cite{cfs_goddard_2023}, from NASA Goddard Space Flight Center, is a versatile software framework that facilitates integrating robust DTN capabilities tailored for space missions.
    % Unibo
    \item \textbf{Unibo DTN}~\cite{caini2023unibo}, from Bologna University, stands out for its focus on academic research, providing a fresh perspective on DTN protocols.
    % DTN7
    \item \textbf{DTN7}~\cite{dtn7} is a flexible open-source implementation of BPv7 that provides modular interfaces for CLAs, routing, and applications. Variants of it are available in the Go, Rust, and Kotlin programming languages.
    The Rust version was leveraged for demonstrating a browser-based DTN solution \cite{baumgaertner2019bdtn7}.
    % uD3TN
    \item \textbf{µD3TN}~\cite{uD3TN} offers the Consultative Committee for Space Data Systems (CCSDS) Space Packet Protocol (SPP) as well as portability to embedded platforms. Evolved from µPCN\footnote{Releases are available via \url{https://upcn.eu/}.}, µD3TN is the sole protocol stack developed, maintained, and supported by a private company and is the core subject of study in this paper.
    % Postellation, DTN2, Terra are not actively maintained -> excluded. We might exclude IBR as well.
\end{itemize}
Apart from this there are a few DTN protocol implementations lacking maintenance (such as \textbf{IBR-DTN}~\cite{schildt2011ibr}).

\subsection{From µPCN to µD3TN}

% uPCN
More than ten years ago, initial work on the ``Micro Planetary Communication Network'' (\emph{µPCN})~\cite{impl:upcn} implementation of the DTN Bundle Protocol version 6 (RFC 5050)~\cite{rfc5050} started. 
Shortly afterward, on 29 March 2015, the first public version, v0.1.0, was released under a BSD license. 
The goal was to provide a compact open-source implementation of the Bundle Protocol that can be used on microcontrollers.
In the subsequent releases, µPCN was enhanced with essential features such as support for the Bundle Protocol version 7~\cite{rfc9171} that evolved at that time in the context of the IETF's DTN working group.
Five years after the initial release, in late 2020, the project \emph{µD3TN} was initiated as a fork of µPCN, intending to provide commercial support and integration services. Since then, development has continued in the public repository on GitLab\footnote{\url{https://gitlab.com/d3tn/ud3tn/}}.

% uD3TN
Over the years, the architecture of µD3TN was subject to several fundamental refinements and improvements\footnote{See also: \url{https://gitlab.com/d3tn/ud3tn/-/blob/master/CHANGELOG}}, making the software modular to increase flexibility and compatibility plus improving its performance.
% Contribution 1
In this paper, we document the design decisions made to serve as a reference for other teams working on DTN protocol implementations and stimulate a discussion about potentially enhancing existing and future implementations of the Bundle Protocol.
% Contribution 2
We also describe the latest µD3TN state and compare it with JPL's reference implementation, the ION protocol stack.

% Paper structure
The remainder of this paper is structured as follows. In section \ref{sec:design-decisions}, we provide insights into the evolution of µD3TN (and previously µPCN) in chronological order and reason about the design decisions taken in the course. 
The following section \ref{sec_architecture} documents the architecture of µD3TN. Section \ref{sec_ion} outlines the architecture of the Interplanetary Overlay Network (ION) \cite{ION-DTN} implementation and contrasts it with the one of µD3TN. Finally, in section \ref{sec_outlook}, we give an outlook on future topics and planned developments in the context of µD3TN and the architecture of Bundle Protocol implementations.

\section{Design Decisions and the Evolution of µD3TN}\label{sec:design-decisions}

\subsection{Concurrency and message-passing (2015)}

The first version of µPCN was developed to become an experimental secondary payload on a CubeSat mission leveraging an STM32F4 microcontroller and the FreeRTOS\footnote{\url{https://freertos.org/}} embedded real-time operating system.
The implementation was initiated to evaluate the Ring Road Network concept~\cite{ringroad}. 
For this purpose, a DTN router implementation operating on low-cost hardware was required.
As no suitable implementation of the Bundle Protocol and a deterministic DTN forwarding approach was available for the constrained hardware environment (a port of ION was considered infeasible due to the platform's limitations\footnote{192 KiB RAM, no persistent storage, no POSIX APIs}), it was decided to implement the necessary protocols and algorithms from scratch.

The goal of the high-level architecture was to be internally flexible while providing a small and stable external interface for integration into the overall satellite platform. 
The authors provide a detailed description of the architecture and an evaluation of the resulting first version of the program in~\cite{impl:upcn}.

\begin{figure}[th]
	\centering
	\includegraphics[width=1\linewidth]{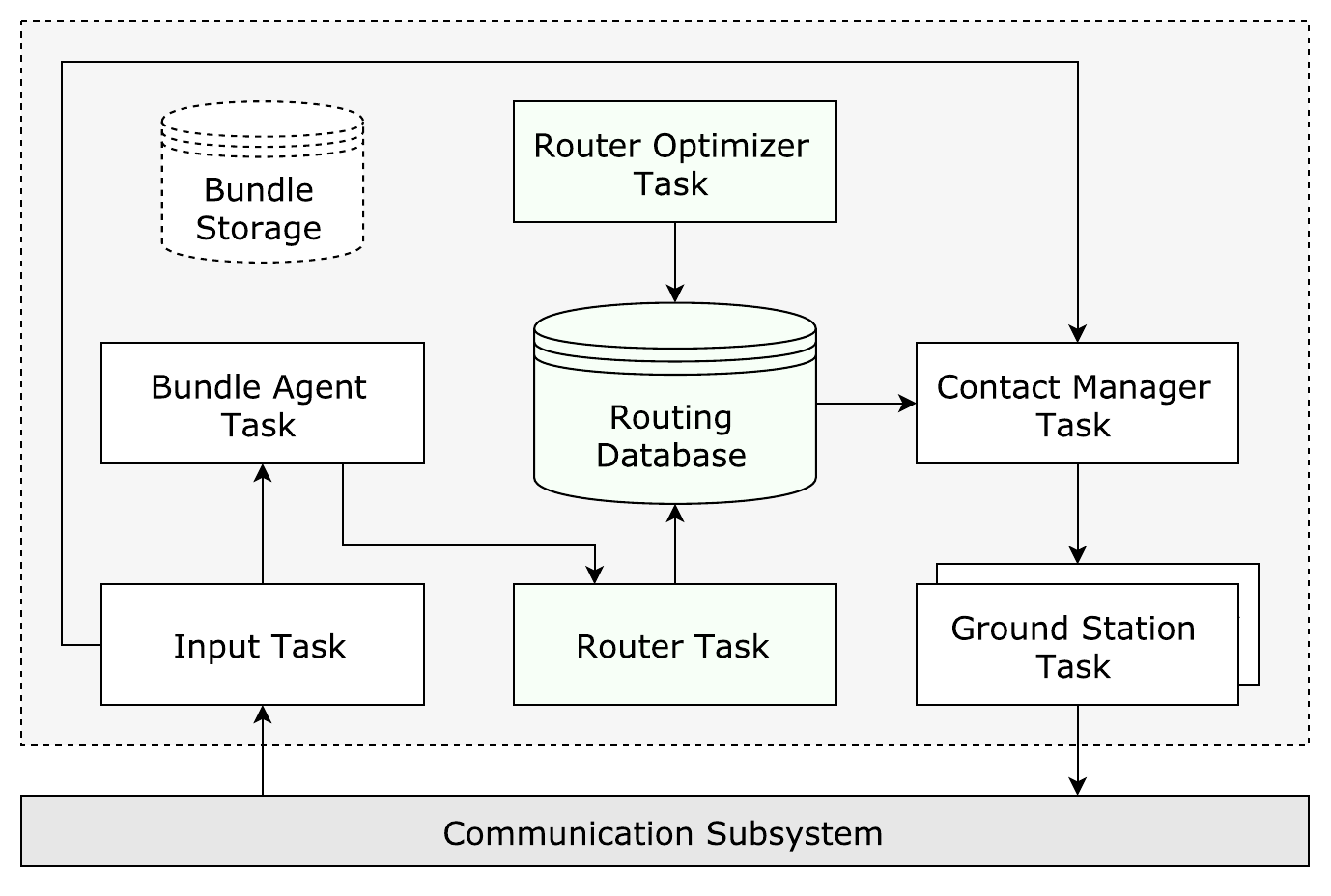}
	\caption{The components of µPCN v0.1.0 as depicted in~\cite{impl:upcn}.}
	\label{fig:upcn-0.1}
	%\vspace{-3ex} % XXX Hack to reduce space between figures
\end{figure}

Figure~\ref{fig:upcn-0.1} gives an overview of the components of µPCN v0.1.0: several independent tasks (depicted as rectangles), connected by message queues, process incoming bundles and configuration commands and perform a simple scheduled next-hop forwarding based on the notion of contacts.

The interface to be leveraged for integration consisted of two message queues (using the queue data structure provided by FreeRTOS), one queue toward the µPCN Input Task for receiving bundles and configuration commands and another queue toward the Communication Subsystem and controlled by the µPCN Ground Station Task. 
The Communication Subsystem was assumed to implement a suitable convergence-layer adapter (CLA). 
Thus, µPCN was implemented independently from any specific CLA.

%By using multiple independent tasks that only interact through message passing, high flexibility concerning managing the scarce resource of processing time on the targeted low-power hardware was achieved: the tasks can receive different scheduling priorities to ensure that less critical operations, such as improving forwarding decisions and contact utilization by reordering bundles, are carried out only when additional processing time is available.
% Splitting in two phrases
By using multiple independent tasks that only interact through message passing, high flexibility concerning managing the scarce resource of processing time on the targeted low-power hardware was achieved.
The tasks can receive different scheduling priorities to ensure that less critical operations, such as improving forwarding decisions and contact utilization by reordering bundles, are carried out only when additional processing time is available.

\subsection{Achieving platform independence (2016)}

While serving the purpose of a technology demonstration and evaluation platform well, the initial versions of µPCN were not very flexible regarding the hardware platform: Only the low-power STM32F4 microcontroller was supported at the beginning, which also requires multiple steps (build, upload, execute) to be performed for testing code changes.
This heavily reduced the development velocity as well as the applicability of µPCN. 
To address this limitation, a port supporting the POSIX API was developed.

POSIX was selected for two reasons:

\begin{enumerate}
    \item There is a wide range of operating systems supporting POSIX; thus, the applicability of µPCN was greatly extended.
    \item It allows for a much more streamlined development workflow, as the code can be written on the same system where it is executed, tested, and debugged, enabling faster iteration cycles.
\end{enumerate}

Besides providing a version for POSIX, it was a central goal not to create a fork of the codebase that needs to be maintained independently of the STM32F4 version. 
Therefore, a \emph{platform abstraction layer} was introduced, which provides an interface to all platform-dependent functionality for which a suitable implementation is selected through the build system.

Due to its compact nature, µPCN requires only a limited number of platform-dependent interfaces that can be implemented efficiently:

\begin{itemize}
    \item an interface for \textbf{task} (thread) creation,
    \item a message-passing \textbf{queue} implementation enabling the interaction between different tasks,
    \item an interface to obtain the current DTN \textbf{time},
    \item a \textbf{semaphore} implementation to control access to shared data structures (e.g., the bundle storage),
    \item an interface to \textbf{input and output} data, and
    \item functions to calculate \textbf{random} numbers and \textbf{checksums}.
\end{itemize}

The released package for µPCN v0.4.0 contains implementations of these functions for both supported platforms and allows for selecting the target platform in the build system.
Enabling several implementations of the platform interface to be developed and maintained next to each other as part of the same code base facilitated efficient maintenance of the software stack while maintaining compatibility, simplifying tests, and simultaneously providing newly developed features and improvements to all target platforms.

\subsection{Concurrent support for BPv6 and BPv7 (2017)}

In parallel with the development of µPCN, the IETF's DTN working group was developing a new version of the Bundle Protocol, and it became clear that future DTN software stacks needed to support it. 
Hence, the authors decided to implement the in-progress draft of the Bundle Protocol version 7 (called \emph{BPbis} at that time) in µPCN. 
This implementation became the first available BPbis implementation as mentioned in the BPbis draft version 7 specification\footnote{See \url{https://datatracker.ietf.org/doc/draft-ietf-dtn-bpbis/07/}, page 36.}.

As discussed in the previous subsection for platform support, the new protocol version was intended to be implemented in the same codebase to be simultaneously maintainable and compatible with both protocol versions.
Unified data structures were created to achieve that, including an in-memory bundle representation that could carry both BPv6 and BPv7 data. 
Although the wire representation of a bundle differs significantly between BPv6 and BPv7, there is a significant overlap concerning the resulting in-memory data structures, making a unified representation possible.

% Idea: sketch how the BPv7 implementation was performed, which libraries were selected, and how, and reference the student thesis!

The version of incoming bundles is distinguished based on their headers—the first byte of the bundle wire representation uniquely identifies the bundle protocol version.
The determined version is then stored in the internal data structure so that the appropriate serialization routine can be selected when transmitting the given bundle again.
The first version of µPCN with combined support for BPv6 and BPv7 (IETF draft -06) was v0.5.0.

\subsection{An abstract CLA interface (2018)}

Versions of µPCN until v0.5.0 did not provide an integrated convergence-layer implementation but expected the system executing it to provide one. 
For testing purposes, minimal tooling was available, allowing for setting up connections between individual µPCN instances. 
However, to make µPCN compatible with other existing DTN protocol implementations and facilitate integration testing, a state-of-the-art convergence-layer protocol had to be implemented.

An essential feature of the Bundle Protocol is being agnostic of the underlying protocols. 
It foresees the flexible selection of suitable convergence-layer adapters to transport bundles to the next DTN-capable hop.
Consequently, a generic Bundle Protocol implementation should also support multiple CLAs, that can be selected based on the use case.

\begin{figure}[th]
	\centering
	\includegraphics[width=1\linewidth]{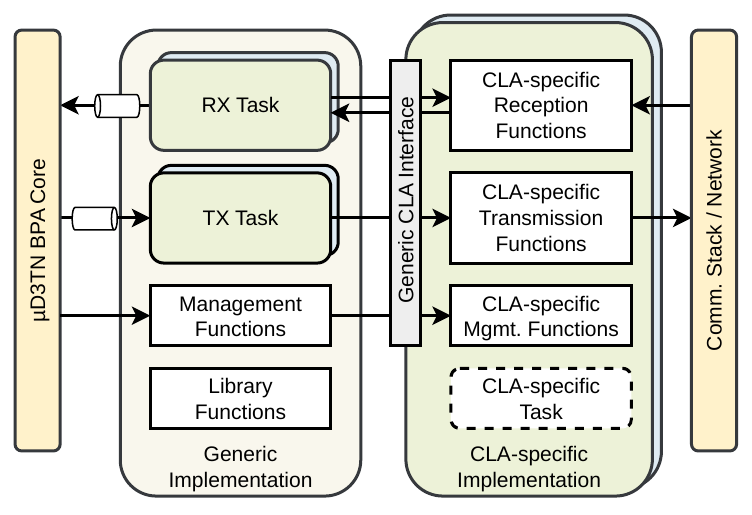}
	\caption{The updated CLA subsystem is separated into a generic part and a CLA-specific part. Both are connected via an abstract interface, which has to be implemented by each particular CLA.}
	\label{fig:cla-impl}
	%\vspace{-3ex} % XXX Hack to reduce space between figures
\end{figure}

In µPCN v0.6.0, the first implementation with multiple CLAs, a similar approach to the platform abstraction layer was used: the CLA had to be selected at compile time and contained a suitable version of all necessary components for sending and receiving bundles.
Due to the inherent code duplication and inflexibility of choosing the CLA at compile time, this concept was quickly abandoned and refactored, resulting in two parts as shown in Figure \ref{fig:cla-impl}; one being generic and one being CLA-specific. Both parts are connected by an abstract, generic interface that is instantiated by each concrete CLA.
This approach allows for building and linking all CLAs supported by the target platform in the same executable and sharing common code.

The generic part of the CLA implementation consists of:

\begin{itemize}
    \item one \textbf{RX Task} per CLA link, handling incoming bundles by invoking CLA-specific functions to receive the wire format and handing it to the appropriate parsers,
    \item one \textbf{TX Task} per CLA link, receiving outgoing bundles through a queue, serializing them, and invoking the CLA-specific functions to send them out,
    \item \textbf{management functions}, which take care of creating and terminating individual CLA and CLA link instances, and
    \item \textbf{library functions} that can be leveraged by a CLA implementation for common tasks such as connection establishment and waiting for data on a socket.
\end{itemize}

The generic part is decoupled from the Bundle Protocol Agent (BPA) through two types of message-passing queues: inbound bundles are put into a shared queue, awaiting BP-compliant processing, whereas one dedicated queue is provided per individual CLA link for outbound bundles.
A \emph{link} in this context can be a dedicated connection using a socket or another form of multiplexing on the CLA level. Links may be constantly available or interrupted and some CLAs may only support a single link in total.
Two tasks are established per link; one instance of the RX Task and one instance of the TX Task. Additionally, CLA implementations may launch further tasks for individual processing or management needs.

As sketched above, each CLA implementation has to instantiate a set of abstract functions, which are invoked by the generic CLA logic. Three groups of CLA-specific functions have to be implemented:

\begin{itemize}
    \item \textbf{reception functions}, which read incoming data and select an appropriate parser for them,
    \item \textbf{transmission functions}, which encapsulate bundles into the desired convergence-layer protocol and send them via an appropriate lower layer, and
    \item \textbf{management functions}, which create and tear down the CLA instance and individual links plus provide status information and general properties of the CLA instance such as the allowed maximum bundle size.
\end{itemize}

Based on this interface, five CL protocols have been implemented as of today: the TCP Convergence-Layer Protocol version 3 (TCPCLv3, RFC~7242), the Minimal TCP Convergence-Layer Protocol (MTCP), a variant of MTCP using a single bi-directional connection, the CCSDS Space Packet Protocol (SPP) proxied over TCP, and Bundle-in-Bundle Encapsulation (BIBE, see section \ref{sec:bibe}).

% NOTE: In the context of this discussion, it could be noted that there is an ongoing effort in the context of the IETF's DTN working group to define the services that a CLA needs to offer ..... % https://www.ietf.org/archive/id/draft-sipos-dtn-bpv7-cla-services-00.html

\subsection{AAP: a flexible application interface (2019)}

Initially focusing on DTN routers in the context of the Ring Road concept, µPCN was only intended to operate nodes forwarding bundles, but not nodes providing bundle endpoints. 
However, it became apparent that allowing the local creation and delivery of bundles enables further use cases and vastly simplifies test scenarios. % when an easy-to-use bundle endpoint is available.

A flexible socket interface was selected for the implementation: the client application can open a connection to µPCN through either a TCP or a POSIX IPC socket. 

For registering bundle endpoints and exchanging the bundle application data, a concise custom protocol with a binary wire format was defined, called the \emph{Application Agent Protocol} (AAP)\footnote{See also: \url{https://gitlab.com/d3tn/ud3tn/-/blob/master/doc/ud3tn_aap.md}}, which supports the following operations:

\begin{itemize}
    \item \emph{REGISTER}: register an endpoint for the provided agent identifier.
    \item \emph{SENDBUNDLE}: send a new bundle containing the provided application data.
    \item \emph{CANCELBUNDLE}: revoke a previous request to send a bundle
    \item \emph{RECVBUNDLE}: sent by µPCN to deliver received bundle application data to the client.
    \item \emph{PING}: check the liveliness of the connection.
\end{itemize}

Every AAP connection registers an own endpoint, which implies that multiple concurrent registrations for the same endpoint are impossible.
Over a single AAP connection, bundle application data units can be exchanged in both directions asynchronously. 
This requires dedicated logic for separating the handling of sending and receiving bundles in applications that need to do both.
However, along with the AAP implementation in µPCN (and now µD3TN), multiple Python tools are provided to quickly integrate AAP functionality in scripts and serve as baseline examples for building more extensive AAP applications.
%The protocol is defined in a specification of the open-source codebase.

\subsection{Enhancements based on OPS-SAT testing (2020-2021)}

In 2020, the latest and last version of µPCN, v0.8.0, was released. 
Shortly afterward, the µD3TN project was initiated and published an initial release v0.9.0. 
Based on this version, multiple field tests of the software were performed in the context of ESA's OPS-SAT mission~\cite{ops-sat}, executing µD3TN onboard the CubeSat as well as on the ground and showing its applicability to realize a Low-Earth Orbit data-ferry network as envisioned by the Ring Road concept~\cite{ringroad}.
These tests moreover provided substantial input to the further development of µD3TN.

The most important findings are:

\begin{enumerate}
    \item Extensive testing with systems ideally identical to the target system is necessary, especially for all interfaces used for integration (such as CLAs). Such tests should be performed early in the development process and, if possible, be integrated into the continuous integration pipeline. We discovered multiple issues during synchronous testing with the OPS-SAT Engineering Model that required ad-hoc solutions and more rounds of testing, specifically related to the communication link leveraged by our SPP CLA.
    \item The implemented contact-based forwarding technique that uses precise timestamps for starting and stopping the emission of bundles provided significant challenges, as the presence of a bidirectional link between satellite and ground at a given timestamp could not be guaranteed. Additionally, the onboard clock and the used time synchronization mechanism were not very precise. 
    \item The core of µD3TN, with its interaction between Bundle Processor, Router Task, and Contact Manager, needed improvements. The tasks require multiple interactions for processing any given bundle, during which, due to shared data structures, they may block each other.
\end{enumerate}

These findings resulted in work to enhance the test toolchain used for validating µD3TN, providing additional scenarios and integrating more platforms on which automatic tests are run.
Because of the issues with the link prediction, a compact contact discovery approach was implemented, which significantly improved the setup's reliability during the field tests.
The tight coupling of the three core tasks in µD3TN was resolved by completely removing the Router Task and making the forwarding decision a synchronous function call.

\begin{figure}[th]
	\centering
	\includegraphics[width=1\linewidth]{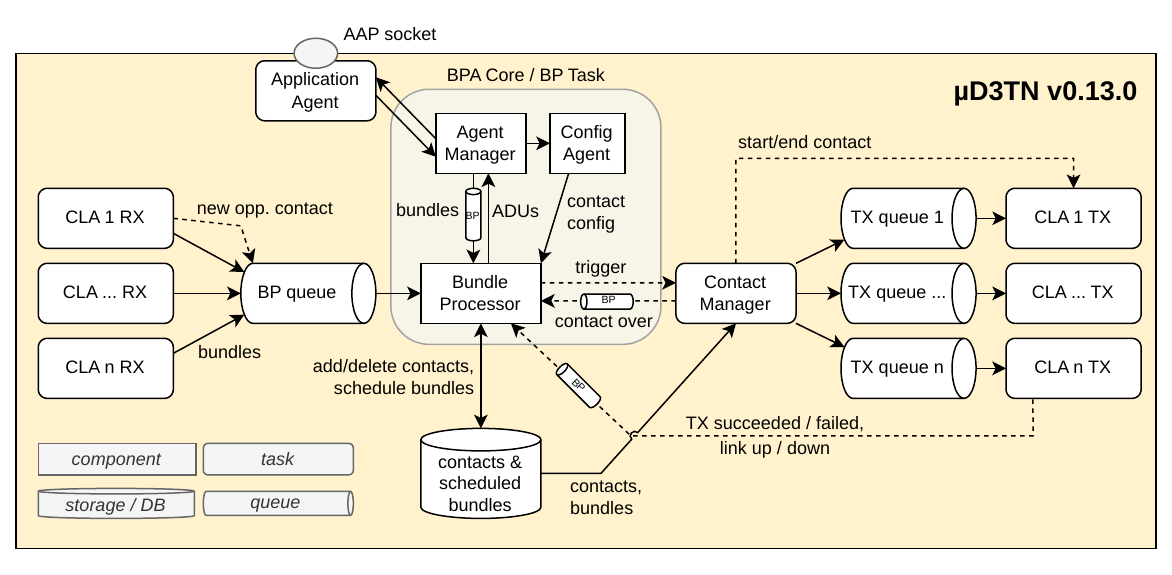}
	\caption{The components of µD3TN v0.13.0 and their interaction, after the removal of the Router Task.}
	\label{fig:ud3tn-0.13}
	%\vspace{-3ex} % XXX Hack to reduce space between figures
\end{figure}

The resulting components and their interactions are depicted in figure~\ref{fig:ud3tn-0.13}.
The findings from the tests and the resulting modifications in µD3TN furthermore triggered discussions on how to flexibly integrate techniques to provide ad-hoc discovery and opportunistic forwarding, which finally led to the \emph{Generic Bundle Forwarding Interface} concept that is discussed below.

\subsection{Bundle-in-Bundle Encapsulation (2021)}\label{sec:bibe}

As the application scope of µD3TN was heavily extended compared to µPCN, now including arbitrary DTN scenarios, work was invested to solve essential network scalability issues when µD3TN is deployed in large-scale networking infrastructures. 
This work was mainly done within the project \emph{REDMARS2}\footnote{\url{https://www.forschung-it-sicherheit-kommunikationssysteme.de/projekte/redmars2}} starting in 2021. 
This project explored concepts from the Recursive Internetwork Architecture (RINA) proposed by John Day in~\cite{rina} for application in the DTN domain. 
Some thoughts about this transfer have been discussed in~\cite{rina-video}.

One of the essential aspects explored in this context was using a generic networking layer to achieve scalability from planetary-scoped networks to ones of interplanetary (or even interstellar) scope. 
The Bundle Protocol was used as this generic layer, applying Bundle-in-Bundle Encapsulation (BIBE)~\cite{bibe} for coupling smaller networks to networks of massive scale, e.g., spanning several planets.

Consequently, the BIBE implementation provided for µD3TN was used within field tests to demonstrate the interaction of nodes connected to the Internet and remote nodes located in an opportunistic DTN based on ad-hoc discovery mechanisms through a deterministic DTN infrastructure that used pre-defined contact plans. 

A core feature of this approach is that it renders encapsulation possible by chaining µD3TN application instances via AAP, up to an arbitrary depth. 
This is enabled by extending the AAP with two messages, \emph{SENDBIBE} and \emph{RECVBIBE}, allowing the AAP client (in this case, a CLA of a higher-layer bundle node) to transmit encapsulated bundles.
This way, locally scoped networks can be coupled via multiple layers of internetworks span on top of them. 
The proposed approach may provide an answer for the scalability issues of broadly scoped, heterogenous DTN infrastructures.

\subsection{Concept: Generic Bundle Forwarding Interface (2022-23)}

\begin{figure}[th]
	\centering
	\includegraphics[width=1\linewidth]{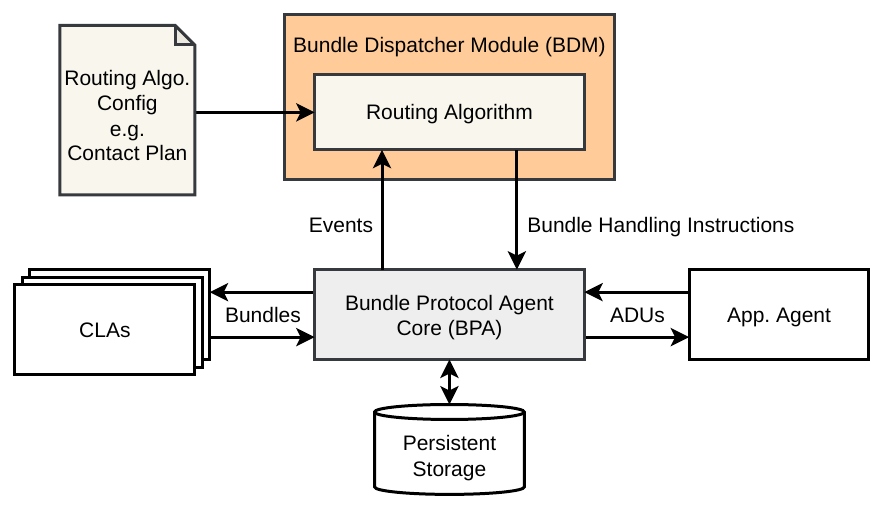}
	\caption{The Generic Bundle Forwarding Interface concept uses a Bundle Dispatcher Module (BDM) for forwarding decisions, as depicted in~\cite{gbfi}.}
	\label{fig:gbfi}
	%\vspace{-3ex} % XXX Hack to reduce space between figures
\end{figure}

%Based on the findings from the OPS-SAT experiments and the REDMARS2 experiments outlined above,
To facilitate operation in heterogeneous networks with varying requirements for routing and forwarding, a concept for modularizing µD3TN's forwarding interface has been developed, which is documented in~\cite{gbfi}. 
The concept introduces a \emph{Bundle Dispatcher Module (BDM)} as shown in Figure~\ref{fig:gbfi}: events concerning the flow of bundles through the Bundle Protocol Agent trigger this external module, which ultimately decides the actions to take for every bundle, such as forwarding it to a specific next hop, persisting it in storage until further notice, or dropping it.

The outlined concept was evaluated in the REDMARS2 project in the context of an experimental port of µD3TN to the Rust programming language (named \emph{µD3TN-NG}).
As µD3TN is implemented in C, it occasionally encounters issues related to memory safety.
Although extensive unit, integration, and fuzz testing\footnote{\emph{Fuzz testing}, also called \emph{fuzzing}, refers to automated testing for issues such as crashes (e.g., in parsers) by feeding the program with a considerable number of randomly crafted, typically unexpected, inputs.} are performed, the port to a memory-safe language was seen as a more sustainable solution and, thus, was tackled in concert with the restructuring efforts.

The overall proof-of-concept implementation was successfully deployed in the REDMARS2 UAV field test, which confirmed the applicability of the approach to heterogeneous network scenarios and indicated that the flexibility gained with it is highly valuable.
For the experiment, multiple proof-of-concept BDMs were implemented in Python; most importantly, one realizing Contact Graph Routing, one using a static routing table, and one executing the opportunistic approach of Epidemic Routing.
Especially the possibility to adapt and exchange the BDMs at runtime without a restart of the node proved to be beneficial during the experiment.

However, the proof-of-concept implementation exhibited several shortcomings. Most prominently, these issues related to the performance of the BDMs due to limitations of the used Python libraries, and to the fragile interaction between tightly-coupled components written in the Rust and C programming languages.
The latter especially resulted from the port not being a complete re-write, but leveraging important components and data structures still written in C in all parts of the new Rust code.
%Additional drawbacks included that the first interface implementation was highly extensive, adding two communication sockets for remote procedure calls and callback-based publish-subscribe operation beside the existing AAP socket.
An additional drawback was that the proof-of-concept implementation did not include access control for clients, meaning that all processes able to connect to the sockets could control all links and forwarding decisions.
Due to these shortcomings, the implementation was put on hold in favor of a novel interface, which is discussed below.

\subsection{Persistent storage accessible via a CLA (2024)}

Implementing persistent storage in µD3TN had been an open issue for a long time, and it was discussed again in light of the new modular forwarding concept.
The main objectives for the storage feature were 1) that it works well with arbitrary forwarding and routing techniques integrated as BDMs, and 2) that the integration is flexible concerning the storage backend and its implementation to facilitate keeping portability to microcontrollers.

The concept that was finally implemented foresees a special convergence-layer adapter responsible for persistently storing bundles.
This may sound counter-intuitive at first, but it has multiple core advantages:

\begin{itemize}
    \item It does not require changes in the already extensive bundle handling logic,
    \item it makes operating multiple storage backends, even concurrently, straight-forward due to the abstract and modular CLA interface,
    \item it prevents implementing dedicated workflows for serializing and de-serializing bundles for sending them over the network and for storing them on disk,
    \item it removes the need for maintaining persistent state in the bundle processing routines, and
    \item it even allows for delegating storage to another node, just by exchanging the used CLA and next-hop address.
\end{itemize}

From the perspective of the Bundle Protocol Agent, the storage CLA transmits the bundle, with an undefined duration, back to itself in a ``loopback'' fashion.
The ``duration'', i.e., when the bundle is released (or alternatively dropped) from storage, is controlled by the forwarding algorithm.
The latter is defined by the attached BDM, which interacts with the storage CLA via bundles directed at a dedicated endpoint identifier assigned to it.
The interface enables querying, dropping, and recalling bundles from storage and facilitates multiple operations to be executed at once by providing a ``bundle filter'' that can match multiple criteria and bundle header fields including placeholders (wildcards).

Using this overall approach, the core logic for processing bundles inside µD3TN can be kept simple and flexible without needing to maintain persistent state and staying fully independent from the forwarding algorithm(s) used.

\subsection{AAPv2 and forwarding modules (2024)}

The first version of AAP that ships with µD3TN for several years was found to have several limitations:

\begin{enumerate}
    % Issue: Headers/Extensibility
    \item The protocol does not allow encoding bundle headers' contents different from the sender's/receiver's endpoint identifier. This means that other header fields, such as the creation timestamp, processing flags, or the contents of extension blocks, can neither be controlled nor received by an AAP client.
    % TODO: verify this is accurate with Carlo
    For example, this limited the integration with DTNperf~\cite{dtnperf} to being unable to use its window-based congestion control mode, which at least would need the bundle creation timestamp to be provided to the client.
    % Issue: Async Send/Receive
    \item Sending and receiving bundles can occur asynchronously over the same stream socket connection. This requires a careful and more elaborate implementation of AAP client applications that need to send and receive bundles on the same endpoint.
    Using multiple concurrent connections is not possible in this case, as AAP disallows the re-registration of an endpoint identifier already registered for another active connection.
    % Issue: No BDM support
    \item The integration of applications providing further services such as controlling bundle forwarding using the Generic Bundle Forwarding Interface concept (see above) required dedicated interfaces.
    % Issue: Access Control
    \item There is no further access control for configuring the integrated contact-based next-hop forwarding algorithm beyond using AAP. All AAP clients must be trusted, as they can configure arbitrary contacts.
\end{enumerate}

For the above reasons and based on the findings from the REDMARS2 experiments, a new version of the AAP interface (\emph{AAP 2.0}) has been developed, which provides features to address the identified drawbacks:

\begin{figure*}[th]
	\centering
	\includegraphics[width=0.75\linewidth]{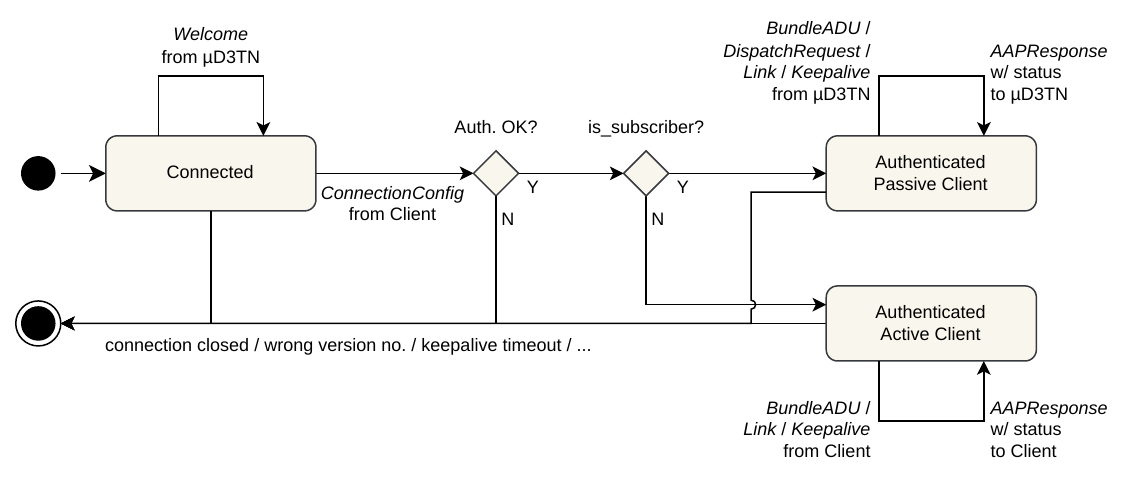}
	\caption{The AAP 2.0 state machine, showing that either the AAP 2.0 client or µD3TN issues commands (remote procedure calls) to the other process, which always have to be answered. The direction of control is set once by the \emph{ConnectionConfig} message and stays that way for the duration of the connection.}
	\label{fig:aap20-statemachine}
	%\vspace{-3ex} % XXX Hack to reduce space between figures
\end{figure*}

\begin{enumerate}
    % Issue: Headers/Extensibility
    \item The new on-wire protocol is specified using the extensible \textbf{Protocol Buffers}\footnote{\url{https://protobuf.dev/}} (Protobuf) format, allowing µD3TN and the AAP 2.0 client to exchange more information concerning bundle headers and enabling extending that set of information in a backward-compatible manner. The Protobuf specification also provides an easily readable, documented interface definition and facilitates straightforward implementation of clients in various programming languages.
    % Issue: Async Send/Receive
    \item \textbf{Sending and receiving bundles is decoupled} in AAP 2.0 and requires two dedicated socket connections. This implies that every AAP 2.0 connection has a fixed direction of control -- either the AAP 2.0 client sends requests to µD3TN or the other way around (for which the client actively switches the direction of control in its initial \emph{CONFIGURE} message.) Figure~\ref{fig:aap20-statemachine} depicts the underlying state machine.
    The issue of potential conflicts between multiple concurrent registrations of the same endpoint is addressed by requiring clients to provide the same value of a ``shared secret'' field, which they can define when establishing the initial connection.
    % Issue: No BDM support
    \item AAP 2.0 gains the capability to \textbf{transmit link updates and bundle forwarding decisions}, enabling BDM operation through a single unified interface.
    % Issue: Access Control
    \item An \textbf{authorization mechanism} is added, requiring applications to request access and provide a secret key for performing operations such as establishing or dropping links and controlling bundle forwarding decisions. 
\end{enumerate}

To realize those new features and adaptations, the AAP 2.0 interface offers the following operations:

\begin{itemize}
    \item \emph{ConnectionConfig}: issued by the client after successfully establishing the connection -- replaces the \emph{REGISTER} operation from AAP version 1, allowing the client to register an endpoint, configure the direction of control, and request authorization for additional actions.
    \item \emph{BundleADU}: transmits a bundle application data unit, i.e., the assembled bundle payload, and relevant metadata to the other peer. Replaces \emph{SENDBUNDLE}, \emph{RECVBUNDLE}, \emph{SENDBIBE}, and \emph{RECVBIBE}.
    \item \emph{DispatchRequest}: asks an authorized client (due to appropriate configuration, a so-called BDM) to calculate a bundle forwarding decision, i.e., a list of next-hop node identifiers plus fragmentation rules to be applied.
    \item \emph{Link}: if issued by an authorized client, requests µD3TN to add a given next hop (including node identifier and CLA address) to its forwarding information base (FIB).
    Based on this, µD3TN will perform necessary actions such as initiating connection attempts through the corresponding CLA.
    This operation can also be used to configure static forwarding rules to skip subsequent calls to the BDM.
    If the control flow has been reversed, \emph{Link} messages are sent by µD3TN to authorized clients to inform them about changes in the FIB (e.g., discovered links or changes performed by other clients).
    \item \emph{Keepalive}: used to check for the liveliness of the connection, whereas a timeout can be configured at compile time. Replaces the \emph{PING} message from AAP version 1.
\end{itemize}

By adding the \emph{DispatchRequest} and \emph{Link} operations, AAP 2.0 provides a unified interface not only for sending and receiving bundles but also for controlling the operation of bundle processing in µD3TN. It thus integrates the \emph{Generic Bundle Forwarding Interface} concept discussed above, enabling the flexible attachment of arbitrary forwarding, routing, discovery, and monitoring clients via a single interface.

%%%%%%%%%%%

\section{Refined Architecture}\label{sec_architecture}

\begin{figure}[th]
	\centering
	\includegraphics[width=1\linewidth]{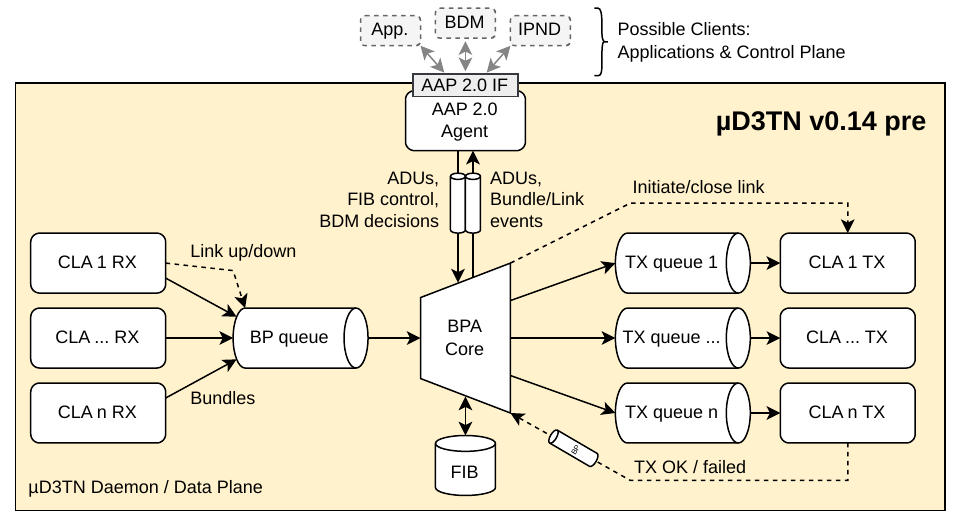}
	\caption{The refined architecture of the µD3TN development tree (mid-2024).}
	\label{fig:ud3tn-1.0}
	%\vspace{-3ex} % XXX Hack to reduce space between figures
\end{figure}

After implementing the improvements discussed in the previous section, the architecture shown in Figure~\ref{fig:ud3tn-1.0} results.

Particularly due to the recent changes in the forwarding subsystem, the core of µD3TN has become a ``bundle multiplexer'' that is controlled by two components:

\begin{itemize}
    \item The \textbf{Forwarding Information Base} (FIB), which associates currently-reachable BP nodes (identified by their DTN node identifiers) with the CLA addresses through which they can be reached. The information held by the FIB is continuously updated when the CLA subsystem reports link changes and when AAP 2.0 clients request links to be established or torn down.
    \item The \textbf{AAP 2.0 Agent} enables forwarding (BDM) and discovery algorithms to be connected to µD3TN, either combined in a single client or provided separately.
    On the one hand, a BDM client is notified by µD3TN about bundles that require a forwarding decision, i.e., bundles for which a FIB lookup did not yield a reachable CLA address. It can then provide a suitable next hop for the bundle, which may be the persistent storage (treated the same as an actual next-hop node due to its implementation as a CLA).
    On the other hand, a discovery client can control the contents of µD3TN's FIB and trigger its CLAs to perform link changes. How the necessary information about reachable nodes is obtained depends on the concrete implementation -- the client may use static configuration, in-band, or out-of-band mechanisms to detect possible neighbors.
    To optimize forwarding performance, the FIB can also be used for ``caching'' next-hop forwarding decisions to prevent triggering the BDM client for multiple subsequent bundles addressed to the same destination node.
\end{itemize}

As a result, the µD3TN daemon only maintains runtime state that can be reconstructed by reloading the BDM, which also obtains and processes information about persistently stored bundles.
The executed forwarding approach can flexibly control the behavior and can also be exchanged at runtime.

The overall refined architecture, which evolved from the initial version of µPCN, allows for comparing µD3TN to \textbf{software-defined networking (SDN)} implementations in state-of-the-art data center networks: A light-weight \emph{data plane} (the µD3TN daemon) is configured by a sophisticated \emph{control plane} that enables external management and can exchange necessary configuration data with its peers.

%%%%%%%%%%%

\section{Comparison to ION Architecture}\label{sec_ion}

The fundamental architectural questions in the design of a DTN implementation may be somewhat clarified by contrasting µD3TN with the Interplanetary Overlay Network (ION) implementation.
Development of ION at NASA’s Jet Propulsion Laboratory began in 2004.  
Over the ensuing two decades, many elements of ION's design have been added, improved, enhanced, and, in some cases, deleted. 
Still, the general architecture was established early on and has proven to be a robust framework for introducing these changes.

\begin{figure*}[th]
	\centering
	\includegraphics[width=0.75\linewidth]{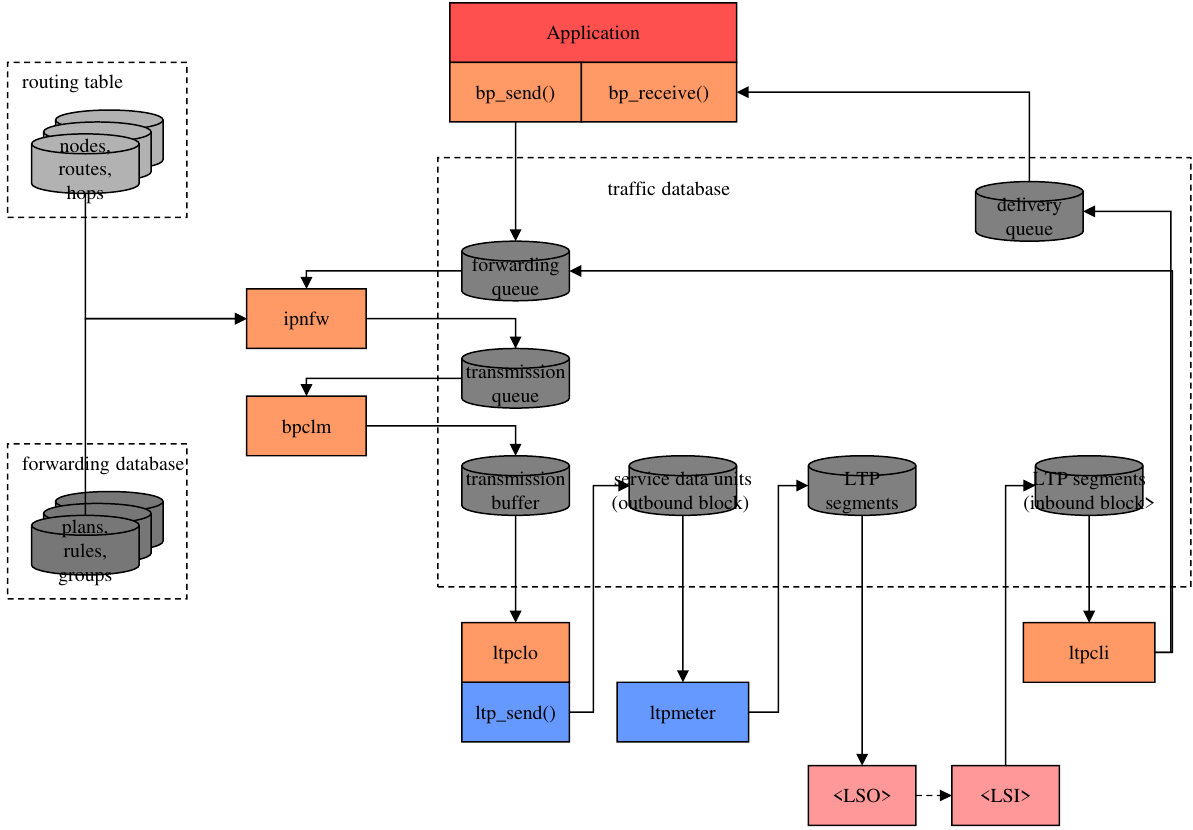} 
	\caption{The ION bundle processing pipeline structure. Taken from the document "\textit{Interplanetary Overlay Network (ION) Design and Operation}", version V4.0.1, dated November 2020 and provided with the ION source code version 4.1.2 downloaded via: \url{https://sourceforge.net/projects/ion-dtn/}.}
	\label{fig:ion}
	%\vspace{-3ex} % XXX Hack to reduce space between figures
\end{figure*}

ION takes the form of a bundle processing pipeline, as illustrated in Figure~\ref{fig:ion}.
%\footnote{Figure \ref{fig:ion} has been taken from the document "\textit{Interplanetary Overlay Network (ION) Design and Operation}", version V4.0.1, dated November 2020 provided with the ION source code version 4.1.2 downloaded via: \url{https://sourceforge.net/projects/ion-dtn/}}.

Bundles and relevant operational data for an ION node are stored in object databases instantiated within notionally persistent shared memory.  
In particular, the ION bundle pipeline comprises several queues in a “traffic” database, each of which is a FIFO-linked list with associated semaphores – one that acts as a mutex, ensuring coherency in the pushing and popping of queue members and one that signals the pushing of new queue members that are available to be popped.

The entry point of the pipeline is the “forwarding” queue, containing bundles for which forwarding is requested.  Bundles are pushed into this queue in one of two ways:

\begin{itemize}
    \item An application may offer new data – an “application data unit” (ADU) – for transmission, causing a new bundle encapsulating that ADU to be created and appended to the queue.
    \item A remotely sourced bundle that is destined for some other node in the network may be received.  The relaying of the bundle is accomplished by appending it to the forwarding queue.
\end{itemize}

A continuously running “forwarder” daemon process pops each bundle of the forwarding queue. 
It determines which node’s “neighbors” (nodes adjacent to this node in the known DTN network topology) should be the next recipient of that bundle on the end-to-end path to its destination.  
To make this determination, the node consults two other databases: a routing table that details the topology of the network and a forwarding database that provides the rules for forwarding bundles through that topology. 
Upon deciding which neighbor should be the bundle’s next recipient, the forwarder daemon pushes the bundle onto that node's “transmission” queue.

The continuously running “convergence-layer manager” (CLM) daemon process corresponding to a given transmission queue pops bundles off the queue and, for each bundle, determines which of several possible “convergence-layer” output (CLO) protocol adapters should be used to forward that bundle.  
Bundle prioritization and rate control are applied at this time, after which the CLM daemon pushes the bundle into the transmission buffer for the selected CLO.

The selected CLO daemon pops each bundle out of its transmission buffer. 
It transmits the bundle to the peer convergence-layer input (CLI) protocol adapter at the selected neighboring node, utilizing the services of the specific convergence-layer protocol for which it was engineered.

ION supports several CL protocols, including TCP, LTP, UDP, DCCP, and others.  
For LTP in particular, an additional layer of pipeline is implemented: the LTP CLO aggregates outbound bundles into LTP “blocks”; an LTP “meter” daemon pops blocks off the blocks queue and fragments them into LTP “segments” for radiation; an LTP link service output (LSO) daemon pops segments off the segments queue and radiates them using one of several possible link service protocols; the peer link service input (LSI) daemon at the selected neighboring node receives the segments and enqueues them for reception by the LTP CLI daemon; and the LTPCLI daemon reconstructs the original LTP blocks and extracts (that is, receives) the bundle(s) encapsulated in those blocks.

Upon receiving a transmitted bundle, the CLI daemon determines whether or not the bundle is destined for the node at which it has been received. 
If so, then the ADU encapsulated in the bundle is extracted and pushed onto the “delivery” queue for the application identified by the bundle’s destination endpoint ID; that application then pops the ADU off the delivery queue and processes it.  
If not, the bundle is appended to the receiving node’s forwarding queue for onward transmission, as noted earlier.

From this overview, we might draw several observations:

\begin{itemize}
    \item Every BP implementation requires a programming interface by which applications can present data units to the BPA for transmission and receive data units transmitted by remote bundle protocol agents.  This interface may be rudimentary as in ION (a simple bp\_send function provided by a library in C) or as extensible as the connection-based Application Agent Protocol supported by µD3TN, depending on the required capabilities.
    \item  A flexible and extensible interface to various convergence-layer protocol adapters will likely be a vital element of any BP implementation. This interface can take various forms, e.g., FIFO queues in ION versus standard functions and message queues in µD3TN.
    \item  It seems to be generally advantageous to factor the bundle processing problem into multiple concurrently running processes that address different discrete aspects. This enables multiple bundles to be processed in parallel (to the extent that the operating system supports this), resulting in a code base comprising multiple relatively simple modules rather than a complex monolith.
    However, allocating functionality to these concurrent processes can be approached in widely differing ways, e.g., the core-centric structure on which µD3TN is built versus the linear pipeline architecture of ION.
    \item  Both µD3TN and ION have had to evolve as support has been required for additional or revised bundle protocol features. Careful attention to fundamental architecture can make that evolution less painful and more successful. During the development process of µD3TN and ION in the last years, their initial architectures have proven significant advantages in this context.
\end{itemize}

%%%%%%%%%%%

\section{Outlook}\label{sec_outlook}

Through a series of architectural refinements over about ten years, the µD3TN Bundle Protocol implementation has grown from an experimental single-purpose application to a flexible and widely applicable professional DTN stack.
Our future work will provide compatible enhancements to the core system and additional capabilities via modules connecting through the AAP 2.0 interface.
In this context, the following concrete contributions are planned in the near future:

\begin{itemize}
    \item A flexible performance testbed is under active development to measure various aspects of the processing and forwarding performance of µD3TN and compare them to multiple other state-of-the-art Bundle Protocol implementations. This effort is expected to lead to substantial performance optimizations and a better overall picture of the available DTN software stacks.
    \item Beside µD3TN, the authors maintain \emph{aiodtnsim}\footnote{\url{https://gitlab.com/d3tn/aiodtnsim}}, a lightweight Python simulation framework for analyzing and rapidly prototyping DTN forwarding and routing techniques. It is planned to enable the use of aiodtnsim forwarding implementations as AAP 2.0 BDMs, which would allow a seamless transition from analyzing new algorithms in simulations to using them to power bundle forwarding through the AAP 2.0 integration.
    \item It is planned to make the CLA interface more flexible and allow dynamic loading/integration of CLAs.
    \item We aim to further pursue the Rust language port of the software to provide a compatible memory-safe alternative to the C implementation.
    \item Work on a proof-of-concept integration with the router platform VyOS\footnote{\url{https://vyos.io/}} is progressing, which may lead to a ``DTN router distribution'' with declarative configuration for straight-forward deployment of DTN nodes.
    \item We plan to leverage more opportunities to test µD3TN in target system contexts to further validate its stability and its flight readiness. Additionally, based on these tests a TRL classification is intended.
\end{itemize}

%%%%%%%%%%%

\section*{Acknowledgments}

We thank all present and past contributors to the µPCN and µD3TN open-source projects.
Parts of the open-source code and multiple concepts presented in this paper were developed in the context of publicly-funded research and development projects that we would like to acknowledge. Specifically:
\begin{itemize}
\item the REDMARS2 project that has been funded by Germany's Federal Ministry of Education and Research (FKZ16KIS1356),
\item the European Union’s Horizon 2020 research and innovation program under the Marie Sklodowska-Curie grant no. 101008233 (``Models in Space Systems: Integration, Operation and Networking''\footnote{\url{https://mission-project.eu/}}), and
\item the D3TN Enterprise Stack (D3TN-ES) project that is co-funded by the European Union.
\end{itemize}

%%%%%%%%%%%

% TODO: Harmonize how we link project web pages -- footnotes (most added by Felix) or bibliography entries (most added by Juan)
\bibliographystyle{IEEEtran}
\bibliography{references}

\end{document}